\input{aipcheck}

\documentclass[
    ,final            
  ]
  {aipproc}

\layoutstyle{8x11single}

\usepackage{graphics}
\usepackage{amssymb,amsmath}
\usepackage{amsmath}
\usepackage{psfrag}
\usepackage{graphicx}
\newcommand{\ve}[1]{\mathbf{#1}}

\newcommand{\f}{\frac}

\newcommand{\be}{\begin{equation}}      
\newcommand{\ee}{\end{equation}}      
      
\newcommand{\bef}{\begin{figure}}      
\newcommand{\eef}{\end{figure}}      
\newcommand{\bea}{\begin{eqnarray}}    
\newcommand{\eea}{\end{eqnarray}}

\newcommand{\bx}{{\bf x}}

\def\spose#1{\hbox to 0pt{#1\hss}}
\def\ltapprox{\mathrel{\spose{\lower 3pt\hbox{$\mathchar"218$}}
\raise 2.0pt\hbox{$\mathchar"13C$}}}
\def\gtapprox{\mathrel{\spose{\lower 3pt\hbox{$\mathchar"218$}}
\raise 2.0pt\hbox{$\mathchar"13E$}}}
\def\inapprox{\mathrel{\spose{\lower 3pt\hbox{$\mathchar"218$}}
\raise 2.0pt\hbox{$\mathchar"232$}}}

\def\bx{{\bf x}}  

\def\br{{\bf r}}

\def\bse{\begin{subequations}}
\def\ese{\end{subequations}}

\def\lsim{\raise 0.4ex\hbox{$<$}\kern -0.8em\lower 0.62ex\hbox{$\sim$}} 
\def\gsim{\raise 0.4ex\hbox{$>$}\kern -0.7em\lower 0.62ex\hbox{$\sim$}}

\def\f0N{f_0^{(N)}}
\def\bec{\begin{center}}
\def\eec{\end{center}}

\begin{document}

\title{Cold uniform spherical collapse revisited}

\classification{95.30.-k, 95.30.Sf, 05.20.-y}
\keywords      {N-body simulations; virialization; spherical collapse}

\author{M. Joyce}{
  address={Laboratoire de Physique Nucl\'eaire et de Hautes Energies,
UMR 7585, Universit\'e Pierre et Marie Curie --- Paris 6, 
75252 Paris Cedex 05, France}
}

\author{B. Marcos}{
  address={Laboratoire J.-A. Dieudonn\'e, UMR 6621, 
Universit\'e de Nice --- Sophia Antipolis,
Parc Valrose 06108 Nice Cedex 02, France}
}

\author{F. Sylos Labini}{
  address={``E. Fermi'' Center, Via Panisperna 89 A, Compendio del
Viminale, I-00184 Rome, Italy}
  ,altaddress={ISC-CNR, Via dei Taurini 19, I-00185 Rome, Italy} 
}


\begin{abstract}
We report results of a study of the Newtonian dynamics of 
$N$ self-gravitating particles which start in a quasi-uniform 
spherical configuration, without initial velocities. These initial
conditions would lead to a density singularity at the
origin at a finite time when  $N \rightarrow \infty$, 
but this singularity is regulated at any finite $N$
(by the associated density fluctuations). While previous studies 
have focussed on the  behaviour as a function of $N$ of the minimal 
size reached during the contracting phase, we examine in particular 
the size and energy 
of the virialized  halo which results. We find the unexpected result that 
the structure decreases in size as $N$ increases, scaling in 
proportion to  $N^{-1/3}$, a behaviour which is associated
with an ejection of kinetic energy during violent relaxation
which grows in propertion to $N^{1/3}$. This latter scaling
may be qualitatively understood, and if it represents the
asymptotic behaviour in $N$ implies that this ejected energy
is unbounded above. We discuss also tests we have performed
which indicate that this ejection is a mean-field phenomenon
(i.e. a result of collisionless dynamics). 
\end{abstract}

\maketitle


\section{Introduction}

We will discuss here numerical simulations, and some simple analytical
results, of what appears at first sight perhaps a very simple initial
condition which one might consider naturally in the study of the 
virialization of self-gravitating systems: a large number $N$  of
point particles are randomly distributed in a sphere, and evolved
from a cold (i.e. zero velocity) start. The ``simplicity'' of the 
family of initial conditions is that they are characterized by
a single parameter, the particle number $N$ (as the system is open 
the units may be defined by the system size, its total mass and 
Newton's constant $G$). Further as $N \rightarrow \infty$ the
initial conditions tends to that considered as a starting point 
for analyses of non-linear structure formation in cosmology
--- the ``cold spherical collapse'' model, in which a perfectly
uniform spherical overdensity embedded in an expanding universe
is considered. A random sampling with particles of such a flat
density profile is the simplest discrete realisation of this 
theoretical model one can envisage, and might thus be expected
perhaps to be the subject of much study. In practice this
initial condition has been but little studied --- references
to the few previous studies will be given below ---  because of the 
intrinsic difficulty in its numerical integration which is
related to a property of the uniform limit: as $N \rightarrow \infty$ 
the evolution of the system leads to a singularity after a finite
time, as all the mass arrives at the origin after a time
$t_{coll} \sim 1/\sqrt{G \rho_0}$ where $\rho_O$ is the initial
mass density. While for any finite $N$ system the singularity
does not occur, the typical size of the region the system contracts
to before ``turn-around'' decreases as $N$ increases, and the
typical particle velocities grow. This makes the numerical
integration very costly, and limits greatly the accessible 
particle $N$ compared to other warm or less homogeneous 
initial conditions.

One motivation for this study comes thus from the ``uniform
spherical collapse model'': when calculating predictions for
the masses and abundances of halos in the framework, one of
the critical assumptions is that {\it all the mass and energy in the initially
collapsing region is ultimately virialized in the collapsed structure}.
The question arises as to whether this is generically true 
independently of the initial conditions. In the present case
--- which we have noted is, in a simple sense, the ``closest''
initial condition to the exactly uniform case --- it turns out,
interestingly, that this is not a good approximation: as we
will describe, the violence of the collapse leads to
an ejection of energy from the system (as kinetic energy of
particles which escape with positive energy). Our numerical
analysis, coupled to an analytical scaling argument, lead
us to conclude that the ejected energy is in fact {\it unbounded
above} as $N$ increases, so that these collapses can be
characterized as causing ``explosions'', with purely 
Newtonian gravitational physics.    

The initial motivation for our study of these initial
conditions was, however, another one: this class of initial
conditions poses issues about discreteness (i.e. N-dependent,
non-collisional) effects in an interesting way. The aim
of $N$-body simulations in cosmology (and indeed, most such
simulations in astrophysics more generally) is to represent  
the collisionless limit, which means that the results of
such simulations should be explicitly independent of $N$
(or depend very weakly on them at the $N$ considered).
In practice this means that, if the dynamics is indeed
collisionless, it should be possible to define an
appropriate extrapolation of $N$ which gives stable 
macroscopic results.In the present class of initial conditions 
it is clear that the simple extrapolation of $N \rightarrow \infty$ 
described above does not give $N$-independent results --- as
we have already indicated we find explicit $N$ dependences
of, for example, the final size and binding energy of the 
virialized structure. There are then two possibilities:
either the physics of this ejection is not in fact
collisionless --- which one would perhaps imagine might be
the case as the particles do coherently contract into
a very small region --- or else the extrapolation 
in $N$ must be performed in a different way. Through 
a careful study which we will summarize below we have 
shown that it the latter which is the correct explanation.
The physical reason is that the $N$ dependence which 
appears in the final state when one extrapolates ``naively''
is in fact a result of the fact that $N$ controls the 
{\it amplitude of the fluctuations} about uniformity, 
which are the physically relevant quantities. The 
appropriate extrapolation to the collisionless limit 
is that in which the particle number is increased, 
while keeping these fluctuations fixed. This extrapolation 
requires the introduction of an additional characteristic 
scale, which is the scale above which the fluctuations
are kept (approximately) invariant. The existence of
the collisionless limit is, conversely, a result of the
fact that the dynamics is only sensitive  to fluctuations
above some finite scale. The study of this problem for 
these initial conditions thus allow one to understand 
some of the subtleties which may be involved in practically 
testing for collisionality effects in $N$ body systems. This
is in principle an important issue in cosmological 
simulations (see \cite{discreteness3_mjbm} for a detailed discussion).

This proceedings essentially summarizes synthetically results reported 
in much greater detail in \cite{mjbmfsl_halos2008}. Because of
space limitations we refer the reader to this paper for a fuller 
set of references on previous studies of virialization, giving
here only the essential references, in particular to the few 
previous detailed studies of these specific initial 
conditions \cite{aarseth_etal_1988, boily_etal_2002, morikawa_nongauss,
  morikawa_virial}.

\section{ Cold collapse theory}

We recall first the limit $N \rightarrow \infty$.
The radial position $r(t)$ of a test particle in an (idealized) exactly 
uniform spherical distribution of purely self-gravitating matter of initial 
density $\rho_0$ and initially at rest (at time $t=0$) is simply
given by the homologous rescaling 
\be
r(t) = R(t) r(0)
\ee
where the {\it scale factor} $R(t)$ may be written in 
the standard parametric form
\bea
&& 
R(\xi) = \frac{1}{2} (1 + \cos(\xi))
\\ \nonumber 
&&
t(\xi) 
= \frac{\tau_{scm}}{\pi} \left( \xi + \sin(\xi) \right)
\;, 
\label{scm2}
\eea 
with 
$\tau_{scm} \equiv \sqrt{\frac{3\pi}{32 G \rho_0}}$.
At the time $\tau_{scm}$ 
the system collapses to a singularity, and physical quantities
diverging. More specifically, taking $\xi = \pi -\epsilon$ 
and expanding to leading order in $\epsilon$ gives
$(t-\tau_{scm}) \sim \epsilon^3$, we have that 
$R(t) \sim [t-\tau_{scm}]^{2/3}$  and therefore the test 
particle velocities $v(t)$, proportional also to the 
initial radius $r(0)$, scale as $v(t) \sim [t-\tau_{scm}]^{-1/3}$

$N$ randomly placed particles in a spherical volume 
can be treated, up to some time and at sufficiently large scales,
as a perturbed version of this uniform limit. 
Consider the approximation in which we treat the perturbations 
as if they evolve also in an infinite contracting system
(i.e. neglect the effect of the boundaries on the evolution of the
density perturbations).  In the manner standard in
cosmology (for the case of an expanding universe)
one can then consider the fluid limit for the system
and solve the appropriate equations perturbatively
(see e.g. \cite{peebles}). In the eulerian formalism 
this gives, at linear 
order, a simple equation for $\delta (\bx)$, the 
density fluctuation (with respect to the mean density):
\be
\label{deltaevol1}
\ddot \delta + 2 H \dot \delta - 4 \pi  G \rho_0 \delta =0
\ee
where $H(t)=\dot{R}/R$ (dots denotes derivatives with respect to
time) is the contraction (``Hubble'') rate. These equations
are derived in ``comoving'' coordinates $\bx=\br/R(t)$, where
$\br$ are the physical vector positions. Note that  
$R \,\dot{\ve x}\equiv =\dot{\ve r} - \dot R\, \ve x $
i.e., $R(t) \dot{\bx} (t)$ is the ``peculiar'' velocity 
with respect to the ``Hubble flow''.

It is straightforward to show, from  Eq.~(\ref{deltaevol1}),
that, in the limit $R \ll 1$, 
\be
\label{limita0}
\delta(R) \sim R^{-3/2}\,.
\ee
This is simply the usual decaying mode of the expanding 
EdS universe,  which becomes the dominating growing mode in the 
contracting case.

The singular behaviour of the spherical collapse is regulated by
the fluctuations present at any finite $N$ in the initial conditions
we study. A simple estimate of the scale factor $R_{min}$ at which 
one expects the spherical collapse model to break down completely
may be obtained by assuming that this will occur when fluctuations 
at some scale (e.g. of order the size of the system) go non-linear. 
For Poisson distributed particles we have a mass variance
proportional to $N$, and  
so the amplitude of the initial normalized density fluctuations
is proportional to $1/\sqrt{N}$. Using the growth law given in 
Eq.~(\ref{limita0}), we can infer \cite{aarseth_etal_1988, boily_etal_2002}.
\be
\label{collapse-scaling}
R_{min} \propto N^{-1/3} \;.
\ee

\section{Numerical Results}

Details of our numerical simulations, performed using the publicly available 
and widely used GADGET2 code \cite{gadget,gadget_paper}, are 
given in \cite{mjbmfsl_halos2008}. Our study covers a range of $N$ 
between several hundred and several hundred thousand. We mention here 
just one important consideration: instead of the exact
Newtonian potential, the code employs, for numerical reasons,
a two-body potential which is exactly Newtonian above a finite
``smoothing length'' $\varepsilon$, and regularized below this scale
to give a force which is 1) attractive everywhere and 2) vanishes at
zero separation. The (complicated) analytic expression for the
smoothing function may be found in \cite{gadget_paper}. With this
modified force the code does not integrate accurately trajectories in
which particles have close encounters, which lead to very large
accelerations and thus the necessity for very small time steps (which
is numerically costly). However, on the (short) time scales we 
consider such trajectories should not play any significant role
in modifying the {\it macroscopic} properties we are interested
in. The results shown below correspond to a constant value of
$\varepsilon$ in all simulations, a few times smaller than $\ell$ in
the largest $N$ simulation. As we will detail further below when we
discuss the collisionless limit for our system, we have tested our
results in particular for stability when $\varepsilon$ is extrapolated
to {\it smaller} values, and we interpret them to be indicative, on
the relevant time scales, of the $\varepsilon=0$, i.e., the exact
Newtonian limit\footnote{As a test we have also performed simulations using
a code with a direct $N^2$ summation, and without any smoothing.
For the range of $N$ (up to a few thousand) for which we can run
this code over the same physical time-scale, we find excellent
agreement with the results obtained with GADGET2 with the
smoothing we have adopted (see \cite{mjbmfsl_halos2008} for
exact parameter values, as well as details of energy 
conservation etc.).}.

Qualitatively the evolution we observe in all our simulations is the
same, and like that well known in both astrophysics, and, more
generally, in statistical physics for systems with long-range
attractive interactions from sub-virial initial conditions of this
type (i.e. with an initial virial ratio larger than -1) : the
system first contracts and relaxes ``violently'' (i.e. on timescales
of order the dynamical time scale $\tau_{scm}$) to give a virialized,
macroscopically stationary, state (see e.g. \cite{binney}).

\subsection{Minimal size: phenomenology}

The existing studies in the astrophysical literature of this class
of initial conditions \cite{aarseth_etal_1988, boily_etal_2002}
focus on how the singular collapse of the uniform spherical
collapse model is regulated at finite $N$, and in particular on
the scaling with $N$ in numerical simulations of the minimal 
size reached by the system. Indeed in the study 
of  \cite{aarseth_etal_1988} the ``points'' represent 
masses with extension (e.g. proto-stars) and the central question 
the authors wish to address is whether these masses survive
or not the collapse of a cloud of which they are the constituents.  
This minimal radius $R_{min}$ may be defined in different ways,
e.g., as the minimal value reached by the radius, measured from
the center of mass, enclosing $90\%$ of the mass. Alternatively
it can be estimated as the radius inferred from the potential
energy of the particles, the minimal radius corresponding to
the maximal negative potential energy.
The behavior of $R_{min}$, 
determined by the first method, as a function of $N$ is 
shown in Fig.~\ref{fig_rmin}. The fitted line 
$R_{min} \propto N^{-1/3}$ is the theoretical behavior 
predicted by the simple arguments given in the previous 
section. Agreement with this simple prediction has
also been verified in both \cite{aarseth_etal_1988} and 
\cite{boily_etal_2002}, the latter for an $N$ as large
as $10^7$. 

\bef
\centerline{\includegraphics*[width=8cm]{FIG1.eps}}
\caption{ Behaviour of the minimal radius $R_{min}$ attained,
determined as described in text,  as a function of $N$.
The solid line is the best fit to the prediction 
of Eq.~(\ref{collapse-scaling}).
\label{fig_rmin}
}
\eef

\subsection{Mass ejection}

All particles start with a negative energy, but a finite fraction 
can in principle end up with a positive energy and escape from
the system completely. While evidently the ejected mass is bounded
above by the initial mass, the ejected energy is, in
principle, unbounded above as the gravitational self
energy of the bounded final mass is unbounded below.

\bef
\centerline{\includegraphics*[width=8cm]{FIG2.eps}}
\caption{ The behavior of $f^p(t)$, the fraction of the particles
with positive energy, as a function of time for two
  different simulations. 
\label{fig_fpt}
}
\eef

\bef
\centerline{\includegraphics*[width=8cm]{FIG3.eps}}
\caption{Behavior of the fraction of ejected particles as a function
of the number $N$ of simulation particles in the system. The solid line
is the phenomenological fit given by Eq.~(\ref{fposlog}).
\label{fig_fpos}
}
\eef

Shown in Fig.~\ref{fig_fpt} is the fraction
$f^p$ of the particles with positive energy as a function of time in
two different simulations, while Fig.~\ref{fig_fpos} shows the 
asymptotic value of $f^p$ in each simulation as a function of 
$N$ (i.e. the value attained on the ``plateau'' in each simulation 
after a few dynamical times, corresponding to particles which 
are definitively ejected on these time scales).While some previous
works (see \cite{mjbmfsl_halos2008} for references) have noted
the ejection of some small fraction of the mass in similar
cases, the significance of the energy ejection as $N$ increases,
and its $N$ dependence, has not previously been documented.
Theoretical studies of the ejection of mass from a pulsating 
spherical system --- which is qualitatively similar to that
described below for the ejection observed here--- can be found in 
\cite{david+theuns_1989, theuns+david_1990}.

Although $f^p$ fluctuates in different realizations with
a given particle number, it shows a very slow, but 
systematic, increase as a function of $N$, varying
from approximately $15\%$ to almost $35\%$ over the
range of $N$ simulated. A reasonably good fit 
is given by  
\be
\label{fposlog}
f^p(N) \approx  a+ b \log(N)\;,
\ee
where $a=0.048$ and $b=0.022$. Alternatively it can be fit 
quite well (in the same range) by a power law 
$f^p \approx 0.1 N^{0.1}$. Note that these fits cannot, 
evidently, be extrapolated to arbitrarily 
large $N$ (as the mass ejected is bounded above), and 
thus our study does not actually definitively determine
the asymptotic large $N$ behavior of this quantity
despite the large particle numbers simulated.
As we will discuss briefly below, however, the 
mechanism we observe for this mass ejection leads 
us to expect that the value of $f^p$ should saturate
when $f^p \sim 0.5$.

\subsection{Energy ejection}

Let us now consider the energy carried away by these 
particles. Using  total energy conservation,
and the fact that both the final potential
energy of the ejected particles and that
associated with the interaction of the
bound and ejected particles is negligible,
we have
\be E_0 = W^n + K^p + K^n \;.
\ee 
where $E_0$ is the initial energy of the system,
$W^n$ and  $K^n$ are the potential and kinetic energy of the particles 
which are finally bound, and $K^p$ is the kinetic energy of
the escaping particles.
Further, since the bound particles in the QSS are virialized
we have
\be
\label{energyapprox2}
2 K^n + W^n =0 \;.
\ee
Thus we have
\be
\label{energyapprox3}
W^n=-2K^n=2(E_0 - K^p) 
\ee

Fig. ~\ref{kposfpos} shows the ratio $K^p/f^p$, i.e.,
the kinetic energy {\it per unit ejected mass}, as a 
function of $N$. Its behavior is fit very well by 
$K^p/f^p \propto N^{1/3}$. Note that for the
largest values of $N$ simulated $K^p$ is
{\it almost ten times} the initial (potential)
energy $E_0$ of the system. It follows that
 we have the approximate behavior 
$W^n \propto -N^{-1/3}$ (when we neglect the slow 
observed variation with $N$ of $f^p$).

\bef
\centerline{\includegraphics*[width=8cm]{FIG4.eps}}
\caption{Behaviour of the total kinetic energy for 
two simulations with different number of particles. 
\label{fig_kinetic_energy}
}
\eef

\bef
\centerline{\includegraphics*[width=8cm]{FIG5.eps}}
\caption{Observed behaviour of the ratio $K^p/f^p$ as a function
of particle number $N$. 
\label{kposfpos}
}
\eef

\subsection{Properties of virialized ``halo''}

The dependence of the ejected energy on $N$ implies that the
macroscopic properties of the virialized structure (which we refer to
as a ``halo'' in the sense current in cosmology)  also depend on $N$. 
Studying the
radial density profiles of the (approximately spherically symmetric)
halos we find that they can always be fit well by the simple functional
form 
\be
\label{dpscm1}
n(r) = \frac{n_0}{\left(1+\left(\frac{r}{r_0}\right)^4\right)} \;.
\ee
This form of the profile agrees well 
with that found in previous studies for collapses from low initial 
virial ratio (see \cite{mjbmfsl_halos2008} for references).
The $N$ dependence we find here is encoded in that of the two 
parameters $n_0$ and
$r_0$, which we find are well fit by $r_0 \propto N^{-1/3}$ and $n_0
\propto N^{2}$.  In Fig.~\ref{fig_dpN} we show the density profiles
for various simulations with different $N$ where the axes have been
rescaled using these behaviors.

It is simple to show that these scalings with $N$ of $n_0$ and $r_0$
are simply those which follow from those just given for $f^p$ and
$W^n$: using the ansatz Eq.~(\ref{dpscm1}) one has that the number
$N^n$ of bound particles is proportional to $n_0 r_0^3$ while the
potential energy $W^n$ is proportional to $m^2n_0^2 r_0^5$ where $m$
is the mass of a particle.  The best fit behaviors for $r_0$ and
$n_0$ thus correspond, since $m\propto 1/N$, to $N^n \sim N$ (i.e. a
constant bound mass, and therefore a constant ejected fraction of the
mass $f^p$) and $W^n \sim N^{1/3}$. More detailed fits to $n_0$ and
$r_0$ show consistency also with the very slow variation of $f^p$
observed.  In summary the $N$ dependence of the virialized structure
manifests itself
to a very good approximation simply in a scaling of its characteristic
size in proportion to $R_{min}$, the minimal radius attained in the
collapse (which, as we have seen, is proportional to the initial
inter-particle separation).

\bef
\centerline{\includegraphics*[width=8cm]{FIG6.eps}}
\caption{Density profile of the virialized structure at a time $t
  \approx 4 \tau_{scm}$ for simulations with different number of
  particles.  The y-axis has been normalized by $N^2$ and the x-axis by
  $N^{-1/3}$ (see text for explanations).  The behaviour of
  Eq.~(\ref{dpscm1}) is shown for comparison. 
\label{fig_dpN}
}
\eef

\section{Further analysis and Discussion}

\subsection{Mechanism of energy ejection}

\begin{figure}
\centerline{
\psfrag{X}[c]{$t$}
\psfrag{Y}[c]{Arbitrary units}
\includegraphics*[width=8cm]{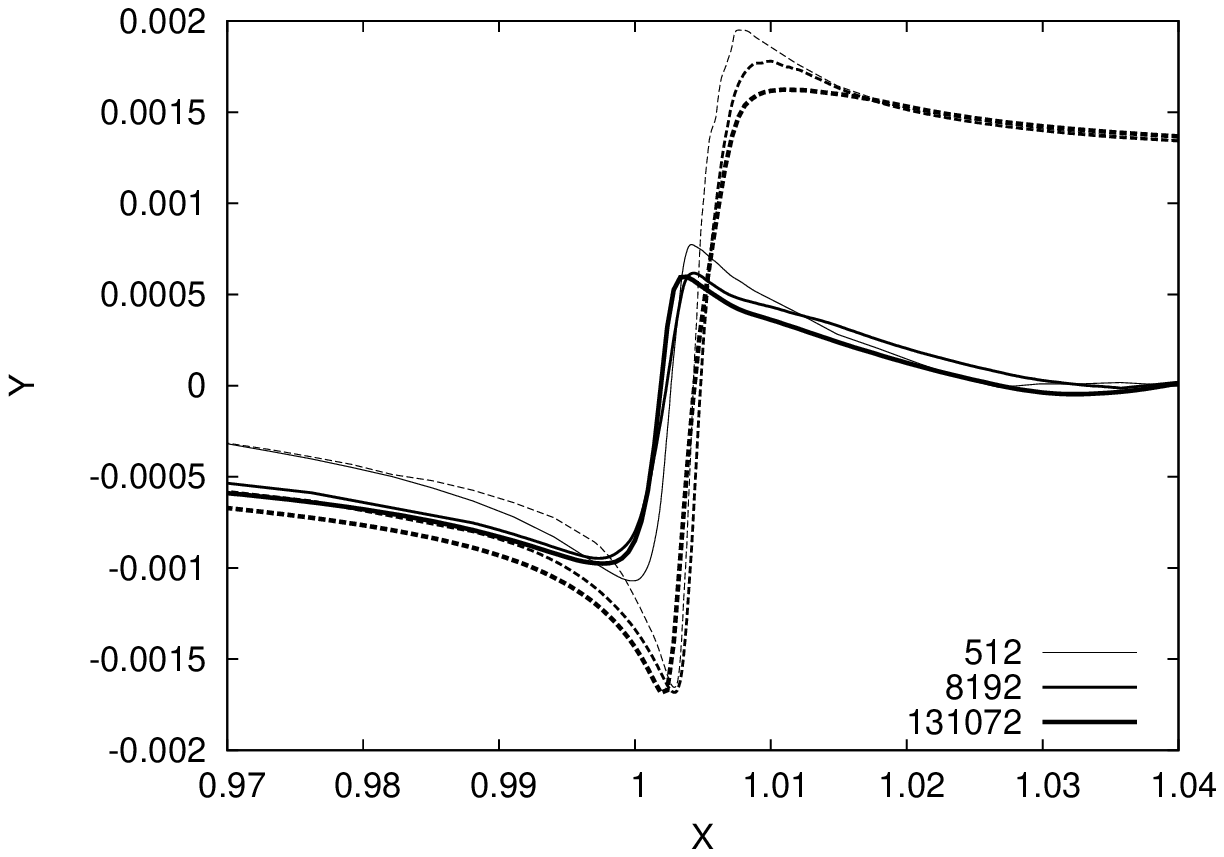}
}
\caption{Radial velocity, and average energy per particle, as a function
of time, of particles which are bound/ejected  at the end of the
simulation of $131072$ particles. The energy of the particles 
has been arbitrarily rescaled.
 \label{radial-131072}}
\end{figure}

We limit ourselves here to a very brief qualitative description of 
a detailed study we have performed (see \cite{mjbmfsl_halos2008}) 
of the evolution of the system during collapse and the
mechanism which leads to the mass and energy ejection.
It is simple to establish that the probability of ejection is closely
correlated with particles' initial radial positions, with essentially
particles initially in the outer shells being ejected. The reason why
this is so can be understood as follows. Firstly,
particles closer to the outer boundary systematically lag (in space
and time) with respect to their uniform spherical collapse
trajectories more than those closer to the center.  This is an effect
which arises from the fact that, when mass moves around due to
fluctuations about uniformity, there is in a radial shell at the
boundary no average inward flux of mass to compensate the average
outward flux.  The mean mass density thus seen by a particle in such a
shell decreases, leading to a slowdown of its fall towards the
origin. This ``lag'' with respect to particles in the inner shells
propagates in from the boundaries with time, leading to a coherent
relative lag of a significant fraction of the mass by the time of maximal
compression. Secondly, these lagging particles are then ejected as
they pick up energy, in a very short time around the collapse, as they
pass through the time-dependent potential of the particles initially
closer to the center, which have already collapsed and ``turned
around''.  This is illustrated in Fig.~\ref{radial-131072}, which
shows, for a simulation with $N=131072$ particles, the temporal
evolution of the components of the mass which are asymptotically
ejected or bound. More specifically the plot shows the evolution of
$v_e$ (and $v_b$) which is the average of the {\it radial} component
of the velocity for the ejected (and bound) particles, and also e$_e$
(and e$_b$) which is the mean energy per ejected (and bound) particle
(i.e. the average of the individual particle energies).  The
behaviors of $v_e$ and $v_b$ show clearly that the ejected particles
are those which arrive on average late at the center of mass, with
$v_e$ reaching its minimum after the bound particles have started
moving outward.  Considering the energies we see that it is in this
short time, in which the former particles pass through the latter,
that they pick up the additional energy which leads to their
ejection. Indeed the increase of $e_e$ sets in just after the change
in sign of $v_b$, i.e., when the bound component has (on average) just
``turned around'' and started moving outward again.  The mechanism of
the gain of energy leading to ejection is simply that the outer
particles, arriving later on average, move through the time dependent
{\it decreasing} mean field potential produced by the re-expanding
inner mass.  Assuming that the fraction of the lagging mass is
independent of $N$, an analysis of the scaling (see 
\cite{mjbmfsl_halos2008})
of the relevant
characteristic velocity/time/length scales allows one to infer the
observed scaling of the ejected energy with $N$. Quantitatively we
have not been able to explain, on the other hand, the observed 
$N$ dependence of the lagging mass, which should determine the $N$ dependence of
$f^p$. Given, however, that it is determined by a lag of the outer
mass relative to that of the inner mass, it seems clear that, as
required, the mechanism observed will naturally lead it to saturate at
a fixed fraction, of order one, as $N$ increases arbitrarily.

\subsection{Is dynamics collisionless?}

The mechanism we have described for the mass and energy ejection is 
clearly of the ``mean-field'' type --- the particles have been considered
to move in the field produced by the bulk of the mass and no role
has made ascribed, notably, to collisions with individual particles.
As the system has contracted so much before re-expanding, particular
care should evidently be taken in verifying that this is indeed the
case, i.e., that the simulations do indeed represent well the 
collisionless limit.  As discussed briefly this is none other than 
a particular case of a question which can be posed about gravitational 
$N$-body simulation, but one which shows the subtleties there may
be in defining an appropriate numerical extrapolation to provide
a clear answer. More precisely the question is whether the system
approximates well the evolution which would be obtained from
a set of ``collisionless Boltzmann equation'' coupled to the Poisson 
equation, i.e. the Vlasov-Poisson (VP) equations. Is this the case? To determine
whether it is we need to understand how we can test for its
validity. As the VP limit is an appropriate $N \rightarrow \infty$
limit for the system, this means specifying precisely how this limit
should be taken. We can then extrapolate our numerical simulations to
larger $N$ to test for the stability of results.

Firstly it is clear that the appropriate extrapolation for the system we have
studied is not the naive limit $N \rightarrow \infty$, i.e., in which
we simply increase the number of Poisson distributed particles: we
have explicitly identified macroscopic $N$ dependencies in fundamental
quantities, so the results at any given $N$ do not approximate those
at any other $N$, and indeed do not converge towards any
$N$-independent behavior.

Formal proofs of the validity of the VP limit \cite{braun+hepp} for a
self-gravitating system require, however, that the singularity in the
gravitational force at zero separation be regulated when the limit $N
\rightarrow \infty$ is taken. This suggests we should take the limit
$N \rightarrow \infty$ while keeping fixed a smoothing scale, like the
$\varepsilon$ we have introduced in our simulations. Doing so we would
indeed expect to obtain a well defined $N$-independent result,
corresponding to the uniform spherical model with such a
regularization of the force: the sphere will not collapse below a
radius of order $\varepsilon$, as the force is then weaker than the
Newtonian force (and goes asymptotically to zero).  One would then
expect to obtain, for sufficiently large $N$, a final state which is
well defined and $N$ independent, but dependent on the scale of the
regularization and indeed even on the details of its implementation.

This limit is not the VP limit relevant here.  We have indeed
introduced a regularization of the force, but this has been done, as
we have discussed, only for reasons of numerical convenience, and our
criterion for our choice of $\varepsilon$ is that it be sufficiently
small so that our numerical results are independent of it.  Our
results are thus, a priori, independent of the scale $\varepsilon$
(and of how the associated regularization is implemented).  To
illustrate this we show in Fig.~\ref{fig_diffeps} the evolution of
$f^p$ (the fraction of particles with positive energy) as a function
of time, in simulations from identical initial conditions with
$N=32768$ particles in which only the value of $\varepsilon$ has been
varied, through the values indicated.  \bef
\centerline{\includegraphics*[width=8cm]{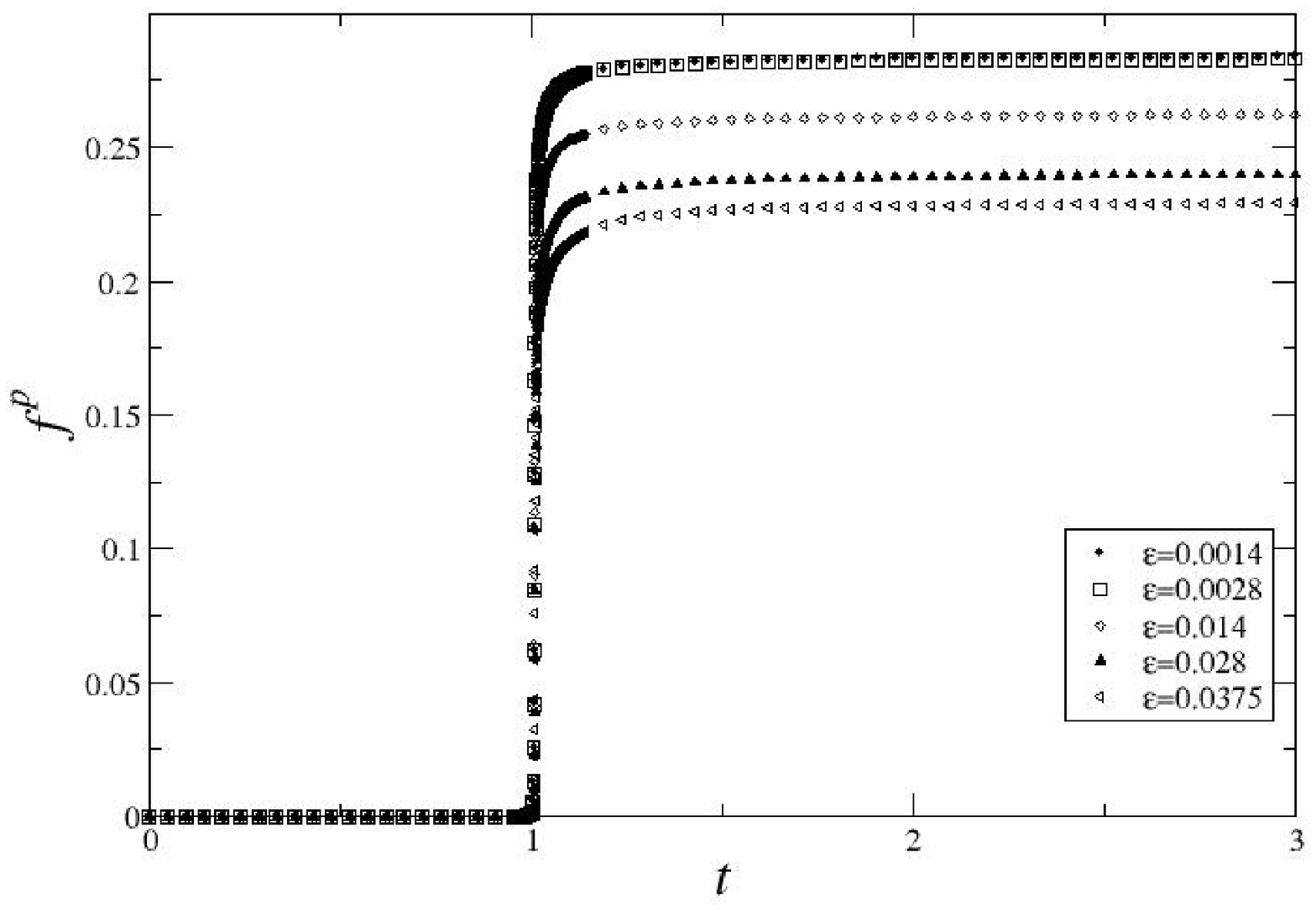}}
\caption{Evolution of the fraction of the mass with positive energy for
simulations with $N=32768$ for the different values indicated
of the smoothing parameter $\varepsilon$.
\label{fig_diffeps}
}
\eef
Other quantities we have considered show equally good convergence
as $\varepsilon$ decreases. Note that for the given simulation
the mean inter-particle distance in our units is $\ell=0.016$, so 
that the convergence of results is attained once
$\varepsilon$ is significantly less than $\ell$.  As we have 
seen, the minimal size reached by the collapsing system scales 
in proportion to $\ell$.  We interpret the observed convergence as 
due to the fact that the evolution of the system is determined primarily
by fluctuations on length scales between this scale
and the size of the system. Once $\varepsilon$ is sufficiently 
small to resolve these length scales at all times, convergence
is obtained.

Given the essential role played by fluctuations to the mean density in
determining the final state, it is clear that only an extrapolation of
$N$ which keeps the fluctuations in the initial conditions fixed can
be expected to leave the macroscopic results invariant. Any change in
$N$ necessarily leads, however, to some change in the fluctuations.
If, however, as indicated by the above results, there is a minimal
length scale in the initial conditions for ``relevant'' fluctuations,
we expect an extrapolation of $N$ which leaves fluctuations above this
minimal scale unchanged to give stable results.

Such an extrapolation for our initial conditions can be defined as
follows. Starting from a given Poissonian initial condition of $N$
particles in a sphere, we create a configuration with $N'=nN$
particles by splitting each particle into $n$ particles in a 
cube of side $2r_{s}$, centered on the original particle. The 
latter particles are distributed randomly 
in the cube, with the additional constraint that
their center of mass is located at the center of the cube, i.e.,
the position of the center of mass is conserved by the ``splitting''. 
In this new point distribution, which has the same mean
{\it mass} density as the original distribution,  fluctuations 
on scales larger than $r_s$ are essentially unchanged compared 
to those in the original distribution, while fluctuations around
and below this scale are modified (see
\cite{gabrielli+joyce_2008} for a detailed study of how fluctuations
are modified by such ``cloud processes''.). We have performed this
experiment for a Poisson initial condition with $N=4096$ particles,
splitting each particle into eight ($n=8$) to obtain an initial
condition with $N'=32768$ particles.  Results are shown in
Fig.~\ref{fig_split} for the ejected mass as a function of time, for a
range of values of the parameter $r_s$, expressed in terms of $\ell$,
the mean inter-particle separation (in the original distribution).
While for $r_s=0.8\ell$ the curve of ejected particles is actually
indistinguishable in the figure from the one for the original
distribution, differences can be seen for the other values, greater
discrepancy becoming evident as $r_s$ increases. This behavior is
clearly consistent with the conjecture that the macroscopic evolution
of the system depends only on initial fluctuations above some scale,
and that this scale is of order the initial inter-particle separation
$\ell$. And, as anticipated, this translates into an $N$ independence
of the results when $N$ is extrapolated in this way for an $r_s$
smaller than this scale.

This prescription for the VP limit can be justified theoretically
using a derivation of this limit through a coarse-graining of the
exact one particle distribution function over a window in phase space
(see e.g. \cite{buchert_dominguez}).  The VP equations are obtained
for the coarse-grained phase space density when terms describing
perturbations in velocity and force below the scale of the
coarse-graining are neglected. A system is thus well described by this
continuum VP limit if the effects of fluctuations below some
sufficiently small scale play no role in the evolution.  The
definition of the limit thus requires explicitly the existence of such
a length scale, and the limit is approached in practice when the mean
inter-particle distance becomes much smaller than this scale. With the
kind of procedure given we have defined not only an extrapolation of
$N$ which gives stable results, but also a method of identifying this
scale.

\bef 
\psfrag{X}[c]{ $t$}
\psfrag{Y}[c]{ $f^p$}
\centerline{\includegraphics*[width=8cm]{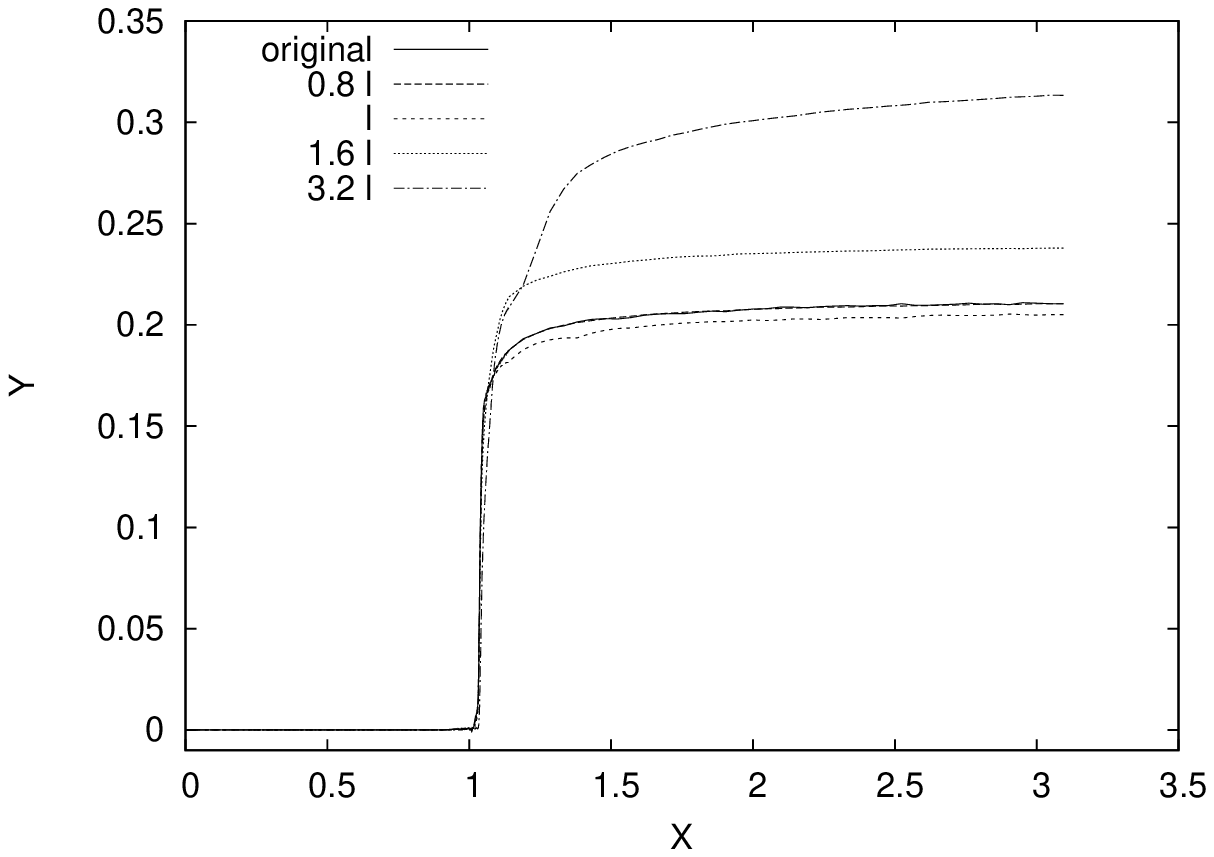}}
\caption{Evolution of the fraction of particles with positive energy 
as a function of time, for the different indicated values of the
parameter $r_s$ described in text. The ``original'' initial conditions
has $4096$ particles while the others have $32768$ particles.
The curve for $r_s=0.8$ is not visible because it is superimposed
on that of the original distribution.
\label{fig_split}}
\eef

\subsection{Some conclusions}

The most surprising result of our study is that a simple initial
configuration of particles interacting by Newtonian gravity only
can liberate, in a time of order the dynamical time, an energy 
which is, apparently, unbounded above, growing approximately 
in proportion to $N^{1/3}$. We have given a physical description
of this ejection which accounts for this scaling, and performed 
careful numerical tests to check that the evolution is indeed
collisionless. This latter step required the definition of the
appropriate numerical extrapolation, which in this case is 
quite non-trivial. This provides an illustration of the subtleties 
which may be involved in determining finite $N$ corrections to the
collisionless limit, which we have argued 
elsewhere  \cite{discreteness3_mjbm} are important notably in 
the context of cosmological simulations of dark matter.

In considering possible astrophysical  relevance of 
these ``classical gravitational explosions'', a question
which evidently arises is whether the energy ejection, which
is associated essentially with the strong contraction during
the collapse phase,  is specific to initial conditions with 
spherical symmetry. In this respect we note that, although we 
expect smaller contraction factors for a generic initial condition, 
very large collapse  factors may occur for non-symmetric initial 
conditions. This question has been considered at length  
in \cite{boily_etal_2002}, which presents a detailed
numerical study of the generalisation of the fluid SCM 
to axisymmetric and triaxial configurations. In
the former singularities remain intact, and a
relation $R_{min} \propto N^{-1/6}$ is
found empirically to replace the $R_{min} \propto N^{-1/3}$ 
behaviour of the spherical case. In the triaxial case 
the collapse factors are found to be typically finite, 
but they can be very large and no upper bound is placed
on them. In forthcoming work we will address the 
ejection of mass and energy also for initial conditions
with non-zero initial velocities. The
divergence we have identified will be regulated again
in this case, but this does not exclude that significant energy 
ejection may occur and be relevant, in particular, to
understanding the properties of the ``remnant'' virialized 
structure.

\begin{theacknowledgments}
We thank the Centro E. Fermi (Rome) for the use of computing resources.
MJ thanks Steen Hansen for useful discussions, and the Istituto dei Sistemi
Complessi, CNR, Rome, for hospitality during several visits.
\end{theacknowledgments}



\bibliographystyle{aipproc}   


\begin{thebibliography}{15}
\expandafter\ifx\csname natexlab\endcsname\relax\def\natexlab#1{#1}\fi
\providecommand{\enquote}[1]{``#1''}
\expandafter\ifx\csname url\endcsname\relax
  \def\url#1{\texttt{#1}}\fi
\expandafter\ifx\csname urlprefix\endcsname\relax\def\urlprefix{URL }\fi
\providecommand{\eprint}[2][]{\url{#2}}

\bibitem[Joyce et~al.(2009{\natexlab{a}})]{discreteness3_mjbm}
M.~Joyce, B.~Marcos, and T.~Baertschiger, \emph{Mon. Not. R. Astron. Soc.}
  \textbf{394}, 751 (2009{\natexlab{a}}), \eprint{arXiv:0805.1457}.

\bibitem[Joyce et~al.(2009{\natexlab{b}})]{mjbmfsl_halos2008}
M.~Joyce, B.~Marcos, and F.~Sylos~Labini, \emph{Mon. Not. R. Astron. Soc}
  \textbf{397}, 775 (2009{\natexlab{b}}).

\bibitem[Aarseth et~al.(1988)]{aarseth_etal_1988}
S.~Aarseth, D.~Lin, and J.~Papaloizou, \emph{Astrophys. J.} \textbf{324},
  288--310 (1988).

\bibitem[Boily et~al.(2002)]{boily_etal_2002}
C.~Boily, E.~Athanassoula, and P.~Kroupa, \emph{Mon. Not. R. Astr. Soc.}
  \textbf{332}, 971--984 (2002).

\bibitem[Iguchi et~al.(2005)]{morikawa_nongauss}
O.~Iguchi, Y.~Sota, T.~Tatekawa, A.~Nakamichi, and M.~Morikawa,
  \emph{Phys.Rev.} \textbf{E71}, 016102 (2005).

\bibitem[Iguchi et~al.(2006)]{morikawa_virial}
O.~Iguchi, Y.~Sota, A.~Nakamichi, and M.~Morikawa, \emph{Phys.Rev.}
  \textbf{E73}, 046112 (2006).

\bibitem[Peebles(1980)]{peebles}
P.~J.~E. Peebles, \emph{{The Large-Scale Structure of the Universe}}, Princeton
  University Press, 1980.

\bibitem[{\textnormal{\texttt{www.mpa-garching.mpg.de/gadget/right.html}}}(200%
0)]{gadget}
{\textnormal{\texttt{www.mpa-garching.mpg.de/gadget/right.html}}} (2000).

\bibitem[Springel et~al.(2001)]{gadget_paper}
V.~Springel, N.~Yoshida, and S.~D.~M. White, \emph{New Astronomy} \textbf{6},
  79--117 (2001), (also available on \cite{gadget}).

\bibitem[Binney and Tremaine(1994)]{binney}
J.~Binney, and S.~Tremaine, \emph{Galactic Dynamics}, Princeton University
  Press, 1994.

\bibitem[David and Theuns(1989)]{david+theuns_1989}
M.~David, and T.~Theuns, \emph{Mon. Not. R. Astr. Soc.} \textbf{240}, 957--974
  (1989).

\bibitem[Theuns and David(1990)]{theuns+david_1990}
T.~Theuns, and M.~David, \emph{Astrophys. Sp. Sci.} \textbf{170}, 276 (1990).

\bibitem[Braun and Hepp(1977)]{braun+hepp}
W.~Braun, and K.~Hepp, \emph{Comm. Math. Phys.} \textbf{56}, 101--113 (1977).

\bibitem[Gabrielli and Joyce(2008)]{gabrielli+joyce_2008}
A.~Gabrielli, and M.~Joyce, \emph{Phys. Rev.} \textbf{E77}, 031139 (2008).

\bibitem[Buchert and Dominguez(2005)]{buchert_dominguez}
T.~Buchert, and A.~Dominguez, \emph{Astron. Astrophys.} \textbf{438}, 443--460
  (2005).

\end{thebibliography}

\IfFileExists{\jobname.bbl}{}
 {\typeout{}
  \typeout{******************************************}
  \typeout{** Please run "bibtex \jobname" to optain}
  \typeout{** the bibliography and then re-run LaTeX}
  \typeout{** twice to fix the references!}
  \typeout{******************************************}
  \typeout{}
 }

\end{document}